\documentclass[dvips]{acta}
\usepackage{supertabular,lscape,epsfig}
\usepackage{amssymb}
\usepackage{amsmath}
\usepackage{multirow}
\usepackage{float}
\mathchardef\mhyphen="2D
\usepackage{placeins}
\DeclareSymbolFont{ppa}{OT1}{ppl}{m}{it}
\DeclareMathSymbol{\vv}{\mathalpha}{ppa}{'166}

\SetPages{103}{112}

\SetVol{71}{2021}

\begin{document}
\newcommand\pvalue{\mathop{p\mhyphen {\rm value}}}
\newcommand{\TabApp}[2]{\begin{center}\parbox[t]{#1}{\centerline{
  {\bf Appendix}}
  \vskip2mm
  \centerline{\small {\spaceskip 2pt plus 1pt minus 1pt T a b l e}
  \refstepcounter{table}\thetable}
  \vskip2mm
  \centerline{\footnotesize #2}}
  \vskip3mm
\end{center}}

\newcommand{\TabCapp}[2]{\begin{center}\parbox[t]{#1}{\centerline{
  \small {\spaceskip 2pt plus 1pt minus 1pt T a b l e}
  \refstepcounter{table}\thetable}
  \vskip2mm
  \centerline{\footnotesize #2}}
  \vskip3mm
\end{center}}

\newcommand{\TTabCap}[3]{\begin{center}\parbox[t]{#1}{\centerline{
  \small {\spaceskip 2pt plus 1pt minus 1pt T a b l e}
  \refstepcounter{table}\thetable}
  \vskip2mm
  \centerline{\footnotesize #2}
  \centerline{\footnotesize #3}}
  \vskip1mm
\end{center}}

\newcommand{\MakeTableH}[4]{\begin{table}[H]\TabCap{#2}{#3}
  \begin{center} \TableFont \begin{tabular}{#1} #4 
  \end{tabular}\end{center}\end{table}}

\newcommand{\MakeTableApp}[4]{\begin{table}[p]\TabApp{#2}{#3}
  \begin{center} \TableFont \begin{tabular}{#1} #4 
  \end{tabular}\end{center}\end{table}}

\newcommand{\MakeTableSepp}[4]{\begin{table}[p]\TabCapp{#2}{#3}
  \begin{center} \TableFont \begin{tabular}{#1} #4 
  \end{tabular}\end{center}\end{table}}

\newcommand{\MakeTableee}[4]{\begin{table}[htb]\TabCapp{#2}{#3}
  \begin{center} \TableFont \begin{tabular}{#1} #4
  \end{tabular}\end{center}\end{table}}

\newcommand{\MakeTablee}[5]{\begin{table}[htb]\TTabCap{#2}{#3}{#4}
  \begin{center} \TableFont \begin{tabular}{#1} #5 
  \end{tabular}\end{center}\end{table}}

\newcommand{\MakeTableHH}[4]{\begin{table}[H]\TabCapp{#2}{#3}
  \begin{center} \TableFont \begin{tabular}{#1} #4 
  \end{tabular}\end{center}\end{table}}

\newfont{\bb}{ptmbi8t at 12pt}
\newfont{\bbb}{cmbxti10}
\newfont{\bbbb}{cmbxti10 at 9pt}
\newcommand{\uprule}{\rule{0pt}{2.5ex}}
\newcommand{\douprule}{\rule[-2ex]{0pt}{4.5ex}}
\newcommand{\dorule}{\rule[-2ex]{0pt}{2ex}}
\def\thefootnote{\fnsymbol{footnote}}
\begin{Titlepage}
\Title{A Survey Length for AGN Variability Studies}
\Author{S.~~K~o~z~{\l}~o~w~s~k~i}{Astronomical Observatory, University
  of Warsaw, Al. Ujazdowskie 4,\\ 00-478 Warszawa, Poland}

\Received{June 25, 2021}
\end{Titlepage}

\Abstract{The damped random walk (DRW) process is one of the most
  commonly used and simplest stochastic models to describe variability
  of active galactic nuclei (AGN). An AGN light curve can be converted
  to just two DRW model parameters -- the signal decorrelation
  timescale $\tau$ and the asymptotic amplitude
  {${S\!F}_{\!\infty}$}. In principle, these two model parameters may
  be correlated with the physical parameters of AGN. By simulation
  means, we have recently shown that in order to measure the
  decorrelation timescale accurately, the experiment or the light
  curve length must be at least 10 times the underlying decorrelation
  timescale. In this paper, we investigate the origin of this
  requirement and find that typical AGN light curves do not
  sufficiently represent the intrinsic stationary process. We simulated
  extremely long (10\,000$\tau$) AGN light curves using DRW, and then
  measured the variance and the mean of short light curves spanning
  1--1000$\tau$. We modeled these light curves with DRW to obtain both
  the signal decorrelation timescale $\tau$ and the asymptotic
  amplitude {${S\!F}_{\!\infty}$}. The variance in light curves
  shorter than $\approx30\tau$ is smaller than that of the input
  process, as estimated by both a simple calculation from the light
  curve and by DRW modeling. This means that while the simulated
  stochastic process is intrinsically stationary, short light curves
  do not adequately represent the stationary process. Since the
  variance and timescale are correlated, underestimated variances in
  short light curves lead to underestimated timescales as compared to
  the input process. It seems, that a simulated AGN light curve does
  not fully represent the underlying DRW process until its length
  reaches even $\approx30$ decorrelation timescales. Modeling short AGN
  light curves with DRW leads to biases in measured parameters of the
  model -- the amplitude being too small and the timescale being too
  short.}{Accretion, accretion disks -- Galaxies: active -- Methods:
  data analysis -- quasars: general}

\Section{Introduction}
Variability of active galactic nuclei (AGN) is of high interest to
astronomers at least for two distinct reasons. The first one is
related to our desire of understanding physics leading to variability
(\eg Kawaguchi \etal 1998, Kelly, Bechtold, and Siemiginowska 2009,
Ross \etal 2018). It seems that the amplitude of light fluctuations
and timescales involved must somehow be related to physics of
accretion disks, to their size, to the accretion flow, and to the
central black hole mass. This is why it is critically important to
accurately characterize an measure the variability of AGN. The second
reason, where our understanding of the variability itself is of less
importance, is to use it as a tool to measure, for example, time lags
in the reverberation mapping method (\eg Peterson 1993). Then AGN
light curves can be modeled as either a stochastic or a deterministic
process -- a Fourier time series, consisting of typically a large
number of basic functions (sines or cosines, \eg Starkey \etal
2016). In such a case the model parameters are generally not the
prime information desired, but a good model describing the light
curve.

We are interested here in accurately measuring variability of AGN
light curves that could be linked to the physical parameters of AGNs.
One of the most widely used models of the last decade has been the
damped random walk (DRW) stochastic process, characterized by just two
model parameters -- the signal decorrelation timescale $\tau$ and the
asymptotic amplitude {${S\!F}_{\!\infty}$} (\eg Kelly, Bechtold, and
Siemiginowska 2009, Koz{\l}owski \etal 2010, MacLeod \etal 2010). While
this model is computationally fast and reproduces AGN light curves
very well (in terms of $\chi^2$), in Koz{\l}owski (2016a) and Koz{\l}owski
(2017b), we presented and explored a number of issues that arise when
modeling AGN light curves with DRW. In Koz{\l}owski (2016b), we showed
that DRW model is able to describe non-DRW stochastic processes
well. This means that we may obtain reasonable DRW fits to data, while
in fact we may not be dealing with the DRW process after all, and
hence the estimated model parameters may be simply meaningless.  In
Koz{\l}owski (2017b), we found that a light curve length must be at least
10 times longer than the decorrelation timescale, otherwise the
measured timescales are underestimated and correlated with the survey
length (the result confirmed by Suberlak, Ivezi{\'c}, and MacLeod
2021). We also showed that measured timescales from currently existing
($\approx$ decade-long) surveys are unlikely to be correct when using
DRW, in particular for AGNs with massive black holes and/or at high
redshifts as the timescale is stretched by $(1+z)$.

In this paper, which is the follow-up paper to Koz{\l}owski (2017b), we
are interested in solving the remaining issue of DRW -- the origin of
the necessity of AGN light curves being longer than 10 times the
decorrelation timescale. We will tackle this problem by simulation
means. In Section~2, we present our simulation setup and the
experiment. In Section~3, we discuss our findings, while in Section~4,
we summarize our results.

\Section{The Experiment}
In this section, we will present our experiment that will lead us to
an answer why the light curves should be 10 (or more) times longer
than the decorrelation timescale. We will describe properties of the
signal, the light curve simulator and the procedure of modeling the
data, a stationary process, and finally we will describe properties of
the experiment.

\subsection{The Covariance Matrix of the Signal}
A light curve is a set of flux measurements taken over a certain time
span.  A relation between two data points ($i$ and $j$) in the DRW
stochastic process is governed by the covariance matrix of the signal
$$S_{\!ij}=\sigma^2 \exp(-|t_i-t_j|/\tau),\eqno(1)$$
for epochs at $t_i$ and $t_j$ (Kelly, Bechtold, and Siemiginowska
2009).  The two model parameters are the signal decorrelation
timescale $\tau$ and the asymptotic amplitude $\sigma$ or
${S\!F}_{\!\infty}=\sqrt{2}\sigma$ (MacLeod \etal 2010). Since the
timescale and the amplitude are correlated, Koz{\l}owski \etal (2010)
introduced a parameter that is less correlated with $\tau$ -- the
modified variability amplitude $\hat{\sigma}=\sigma\sqrt{2/\tau}$.

\subsection{The Light Curve Simulator}
Generating a DRW light curve requires a starting point for the
variable signal that is obtained as $s_1=G(\sigma^2)$, where
$G(\sigma^2)$ is a Gaussian deviate of dispersion $\sigma$. Subsequent
signal values are then iteratively calculated as
$$s_{i+1}=s_i e^{-\Delta t/\tau}+G\left[\sigma^2\left(1-e^{-2\Delta t/\tau}\right)\right],
\eqno(2)$$
where $\Delta t=t_{i+1}-t_i$ (as in Koz{\l}owski \etal (2010), Zu,
Kochanek, and Peterson 2011, Koz{\l}owski 2016b).

\subsection{Modeling the Light Curve}
The full explanation of how to model a light curve with DRW is
presented in Appendix of Koz{\l}owski \etal (2010), Zu, Kochanek, and
Peterson 2011, and Zu \etal 2013, 2016. For completeness, we also
present its basic concepts here.

An AGN light curve ${\bf y}(t)$ may be considered generally as a sum
of the variable signal ${\bf s}(t)$ (having the covariance matrix
$S$), the photometric noise ${\bf n}$ (having the covariance matrix
$N$), and matrix $L$ multiplied by a set of linear coefficients ${\bf
q}$ that are used to subtract or add the mean light curve magnitude
or to remove trends
$${\bf y}(t)={\bf s}(t)+{\bf n}+L{\bf q}.\eqno(3)$$

The likelihood of the data given ${\bf s}(t)$, ${\bf q}$, and model
parameters $\tau$ and $\hat{\sigma}$ is
$$\mathcal{L}\left({\bf y} \bigl| {\bf s},{\bf q},\tau, 
\hat{\sigma}\right)=|C|^{-1/2}|L^T C^{-1} L|^{-1/2}
\exp\left( -\frac{ {\bf y}^T C_\perp^{-1} {\bf y}}{2}\right),\eqno(4)$$
where $C=S+N$ is the total covariance matrix of the data, and
$C_\perp^{-1} = C^{-1}-C^{-1}L (L^T C^{-1} L)^{-1} L^T C^{-1}$. To
measure the model parameters, the likelihood $\mathcal{L}$ is
optimized. To prevent our model from running into unconstrained
parameters, Koz{\l}owski \etal (2010) and MacLeod \etal (2010) used
priors on the likelihood of the model parameters $P(\tau)=1/\tau$ and
$P(\hat{\sigma})=1/\hat{\sigma}$, and we use the same priors here.
Since in this experiment we are not interested in the impact of the
photometric noise ${\bf n}$ on our model parameters, we do not add the
photometric noise to the data. Our light curve ${\bf y}(t)$ is simply
${\bf s}(t)$.

\subsection{Do Light Curves Represent a Stationary Process?}
A stationary process means that all its moments are independent of
time. A weakly stationary process of $N$-th order requires all its
moments up to $N$ to be time invariant. DRW is a weakly stationary
process as it requires its mean, variance, and covariance (and hence
the auto-correlation function, ACF) to be independent of time
(Brockwell and Davis 2002). On the other hand, the Wold decomposition
theorem states that any stationary process can be represented by an
autoregressive process, where DRW is simply a first-order
autoregressive process.

This has led us to a question if a short (as observed by a survey)
light curve was indeed a sufficient representation of the ``full''
intrinsic process. As we will show this is the basic question to the
whole process of light curve modeling with DRW.

\subsection{Simulated Data}
An answer to the above question can be obtained by simulation
means. We decided to simulate long light curves (in terms of $\tau$)
with the length of 10000$\tau$, with the known amplitude
$\sigma=0.2$~mag, and the decorrelation timescale $\tau=300$~d with
two cadences of 3~d (100 points per $\tau$) and 30~d (10 points
per $\tau$). The two light curves are presented in Fig.~1.

\begin{figure}[htb]
\centerline{\includegraphics[width=13cm]{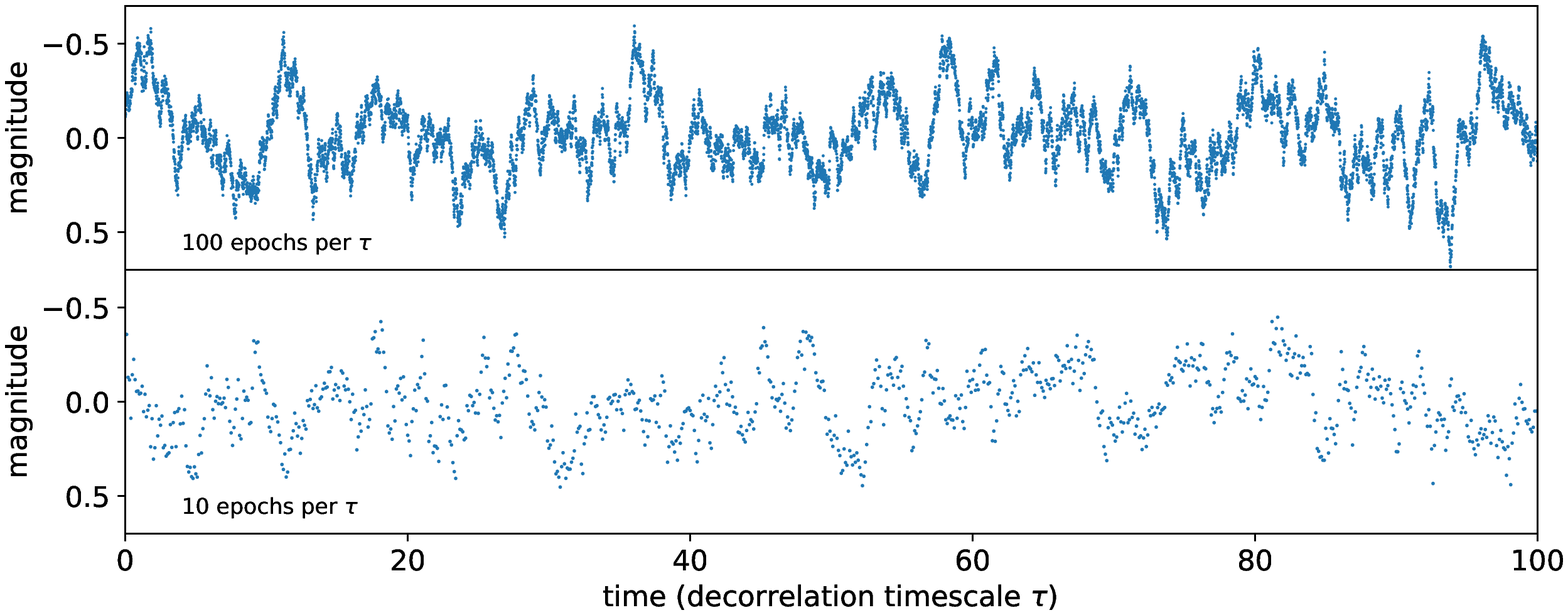}}
\FigCap{Simulated DRW light curves are shown (with $\sigma=0.2$~mag, 
${S\!F}=0.28$~mag, and $\tau=300$~d). They are sampled with 100 epochs 
({\it top panel}) and 10 epochs ({\it bottom panel}) per decorrelation 
timescale $\tau$. While the full simulated light curves span
10000$\tau$, here we present only a small fraction (1\%) of their lengths.}
\end{figure}

\begin{figure}[b]
\centerline{\includegraphics[width=6cm]{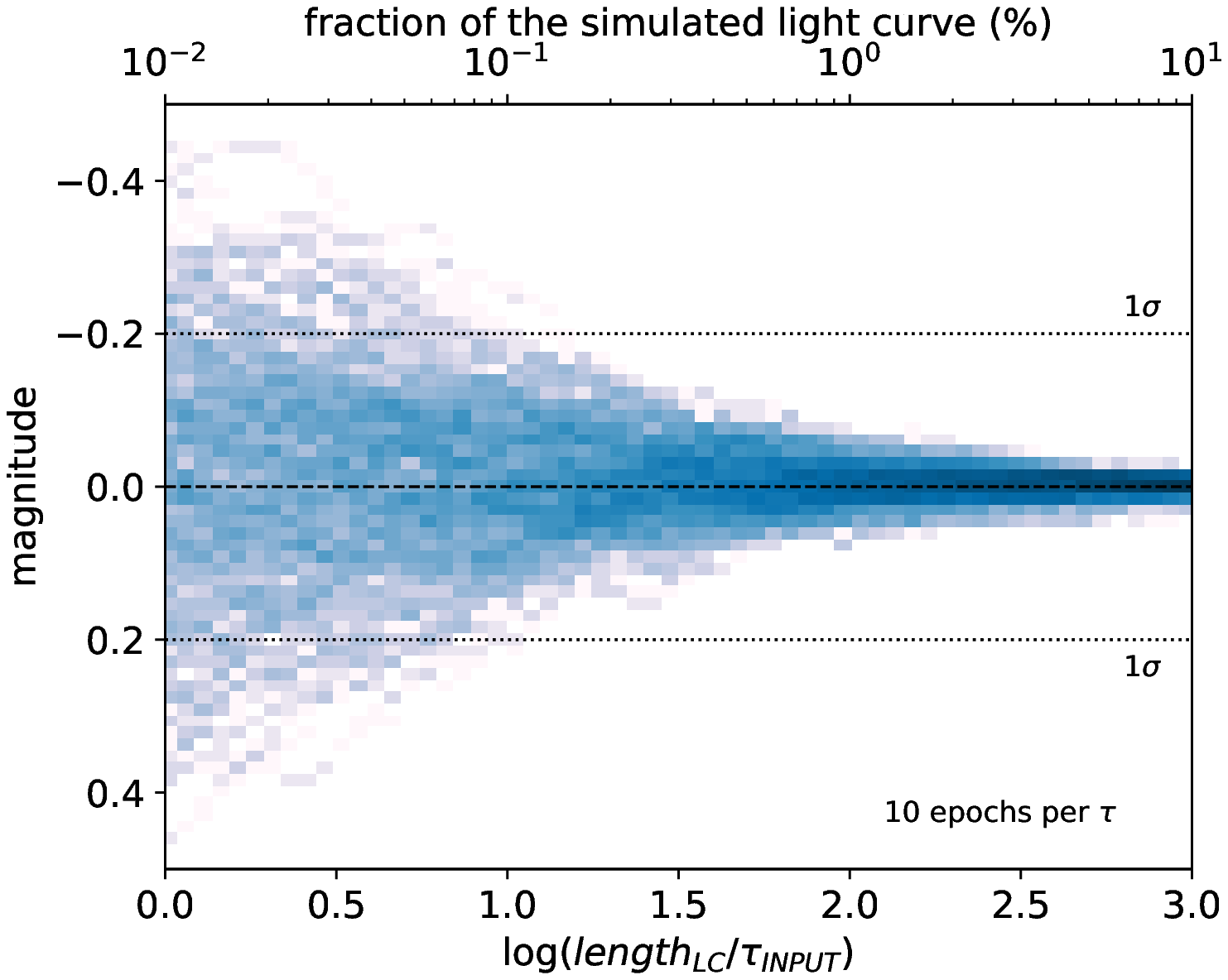}\includegraphics[width=6cm]{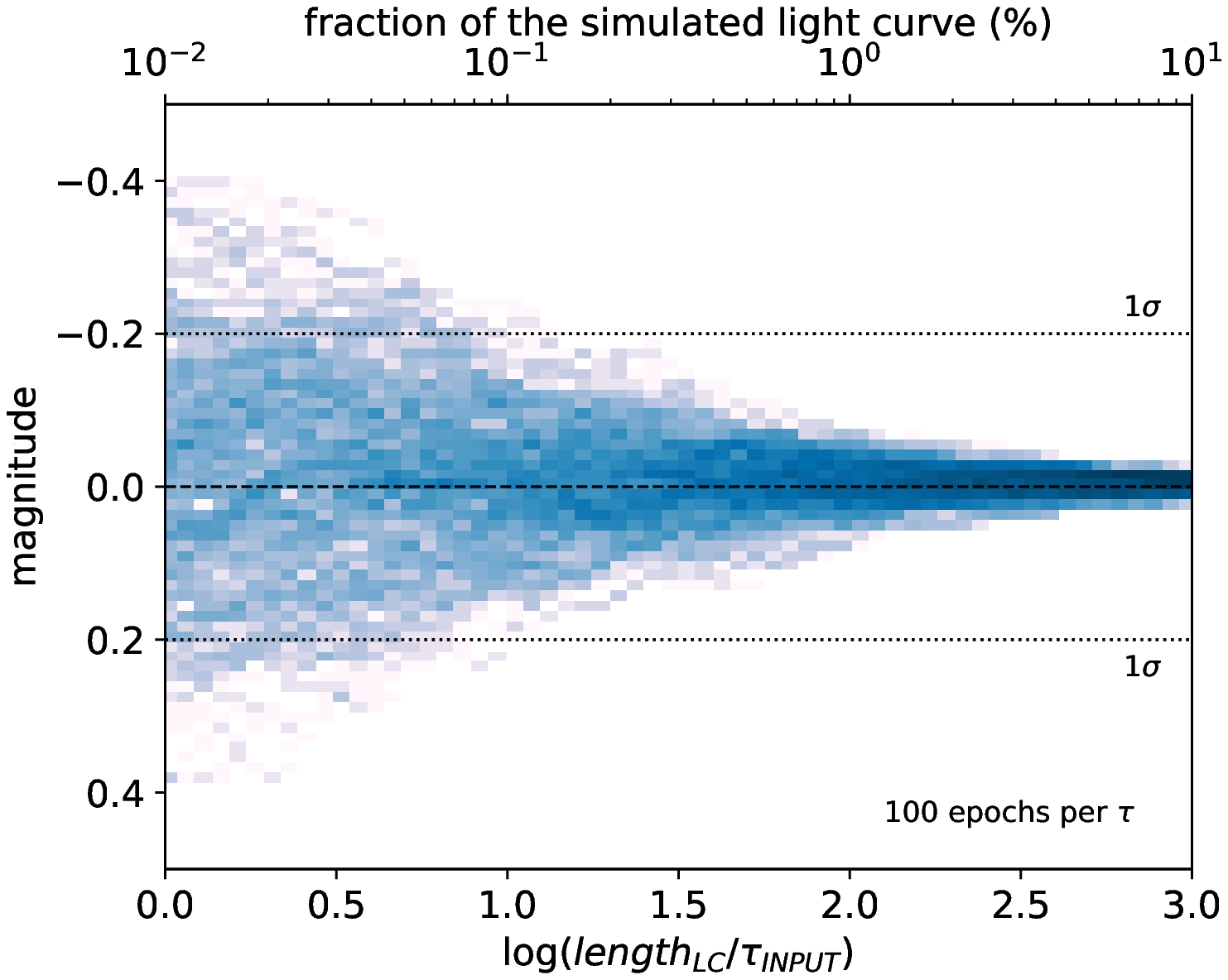}}
\vspace{0.3cm}
\centerline{\includegraphics[width=6cm]{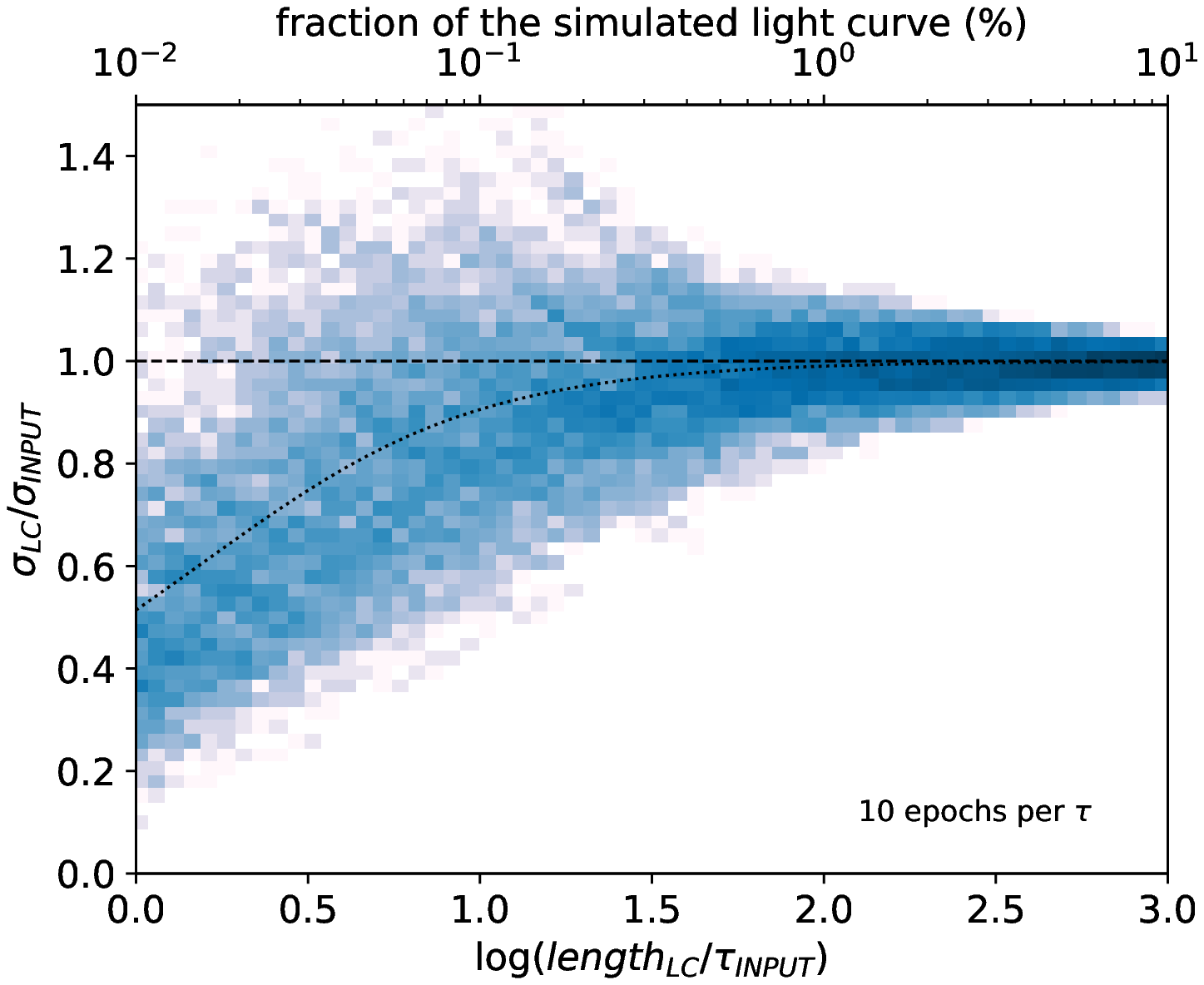}\includegraphics[width=6cm]{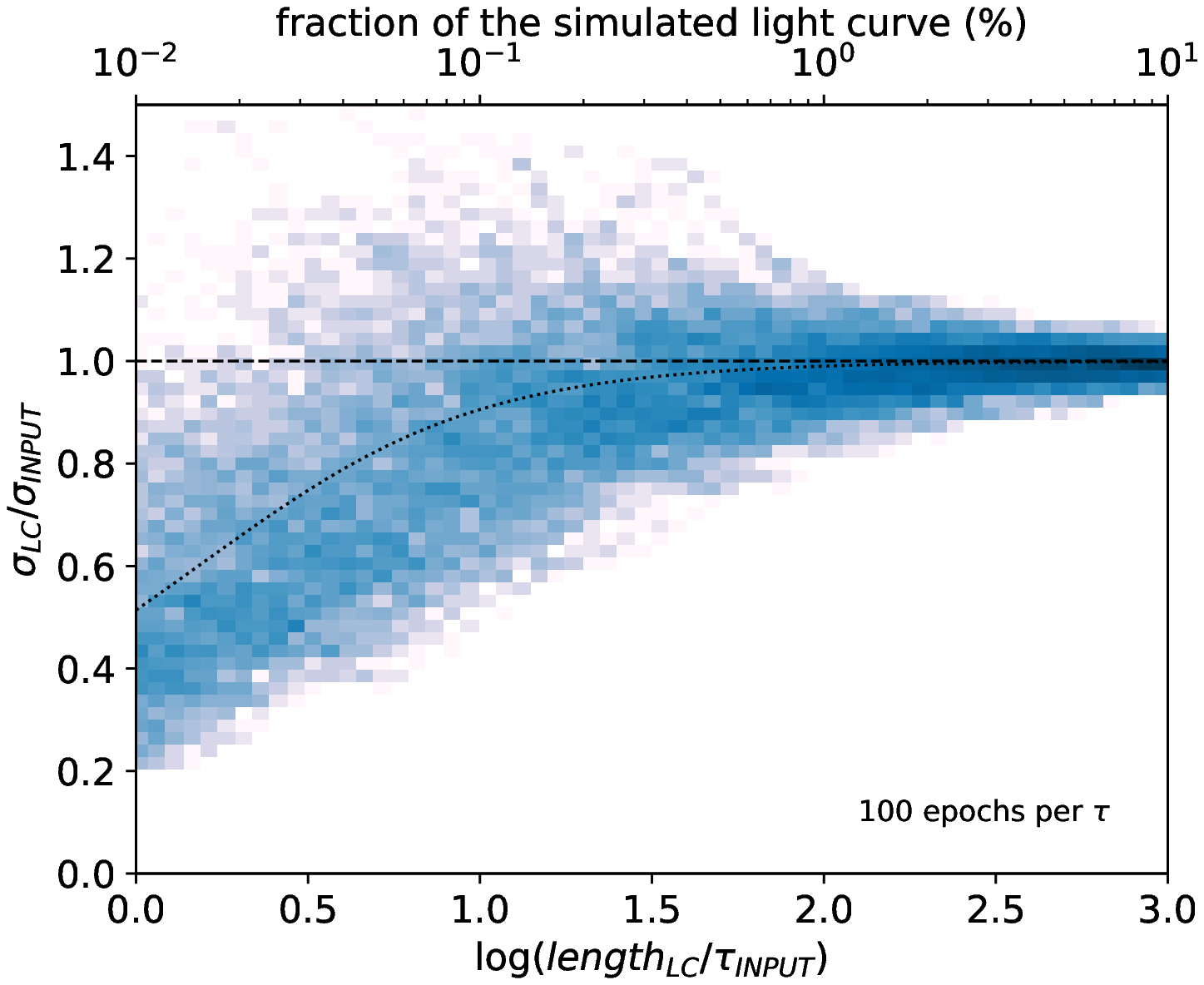}}
\FigCap{The mean magnitude ({\it top panels}) and dispersion ({\it
    bottom panels}) for short light curves as compared to the input
  values (dashed lines). The shorter the light curve, the higher
  the difference between the mean magnitude in the short light curve
  and the input value ({\it top panels}).  The shorter the light
  curve, the smaller the measured dispersion in the light curve as
  compared to the input value ({\it bottom panels}). The dotted line
  shows the theoretical variance from Eq.(5) for the continuous DRW
  process.}
\end{figure}

\Section{Discussion}
We randomly draw short light curves from these two long light curves
with the length between 1--1000$\tau$. These short light curves will
represent observed fractions of the DRW process by astronomical
surveys. For each short light curve, we calculate the mean magnitude
(shown in top panels of Fig.~2) and dispersions (shown in bottom
panels of Fig.~2). From top panels of Fig.~2, we observe that the
shorter the light curve the more is the mean magnitude is deviating
from the input value. In the bottom panels of Fig.~2, we present the
ordinary dispersion as a function of the light curve length. Starting
the inspection of the panels from the right side (\ie the long light
curves), we see that the input and measured dispersions are nearly
identical. Going in the direction of shorter light curves, we observe
that dispersions in light curves becomes increasingly smaller. While
it is difficult to pinpoint the exact moment for this transition it is
safe to say it happens in the range 30-100$\tau$. Both the mean and
dispersion show that light curves shorter than $\approx30~\tau$ no longer
adequately represent the stationary process. This means they do not
have the properties of the stationary process.
\begin{figure}[htb]
\vglue-4mm
\centerline{\includegraphics[width=6cm]{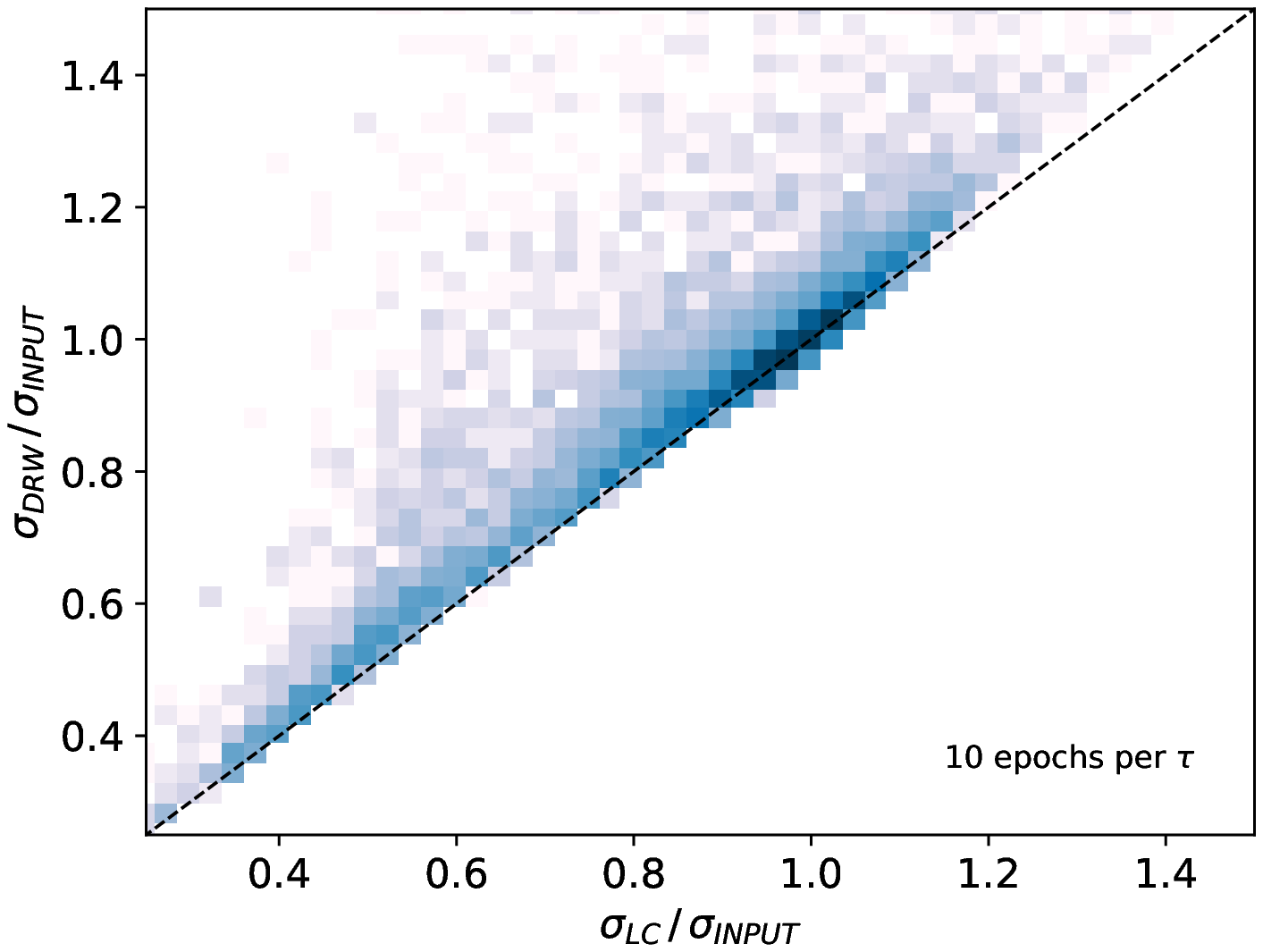}\includegraphics[width=6cm]{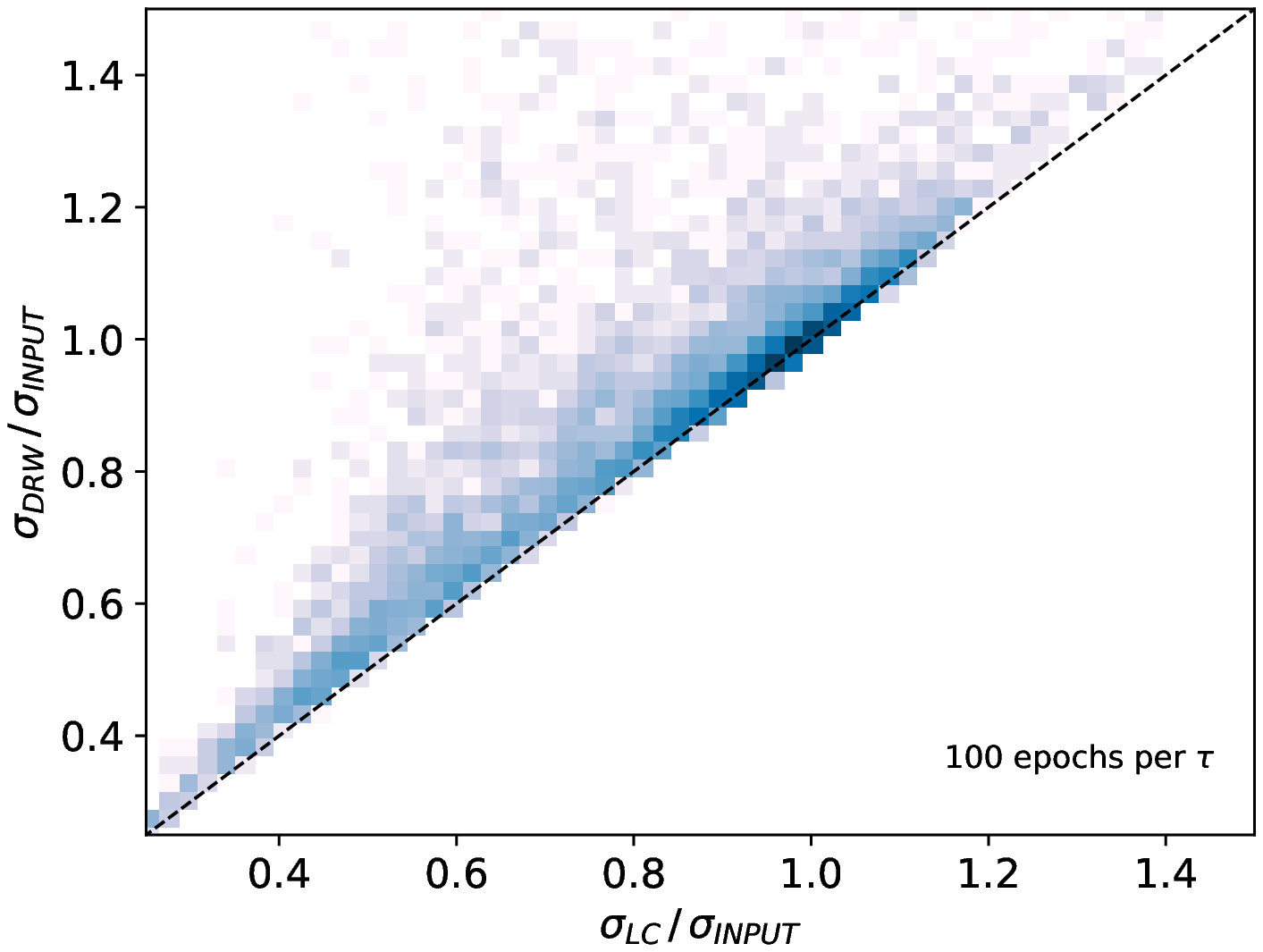}}
\vspace{0.3cm}
\centerline{\includegraphics[width=6cm]{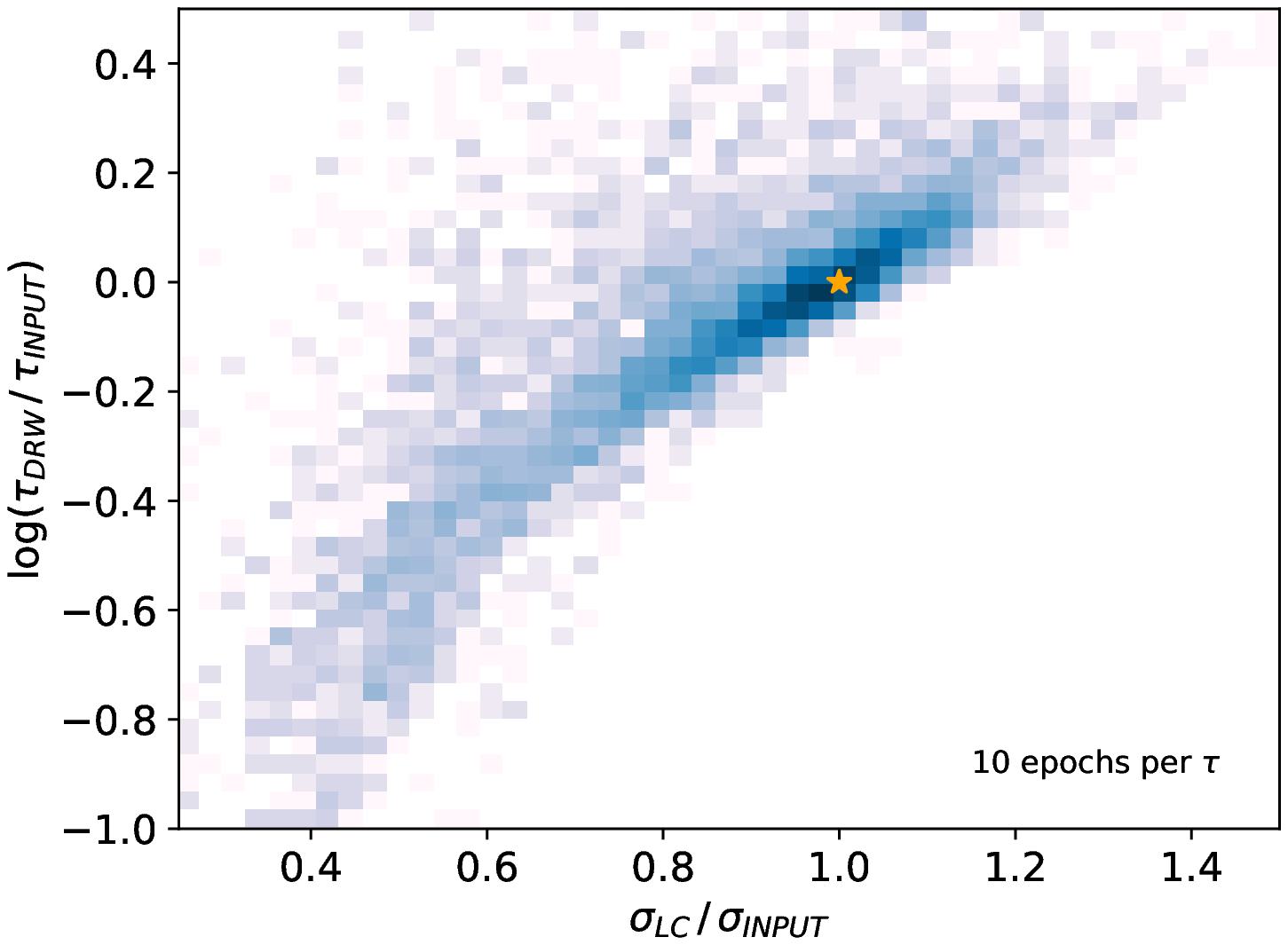}\includegraphics[width=6cm]{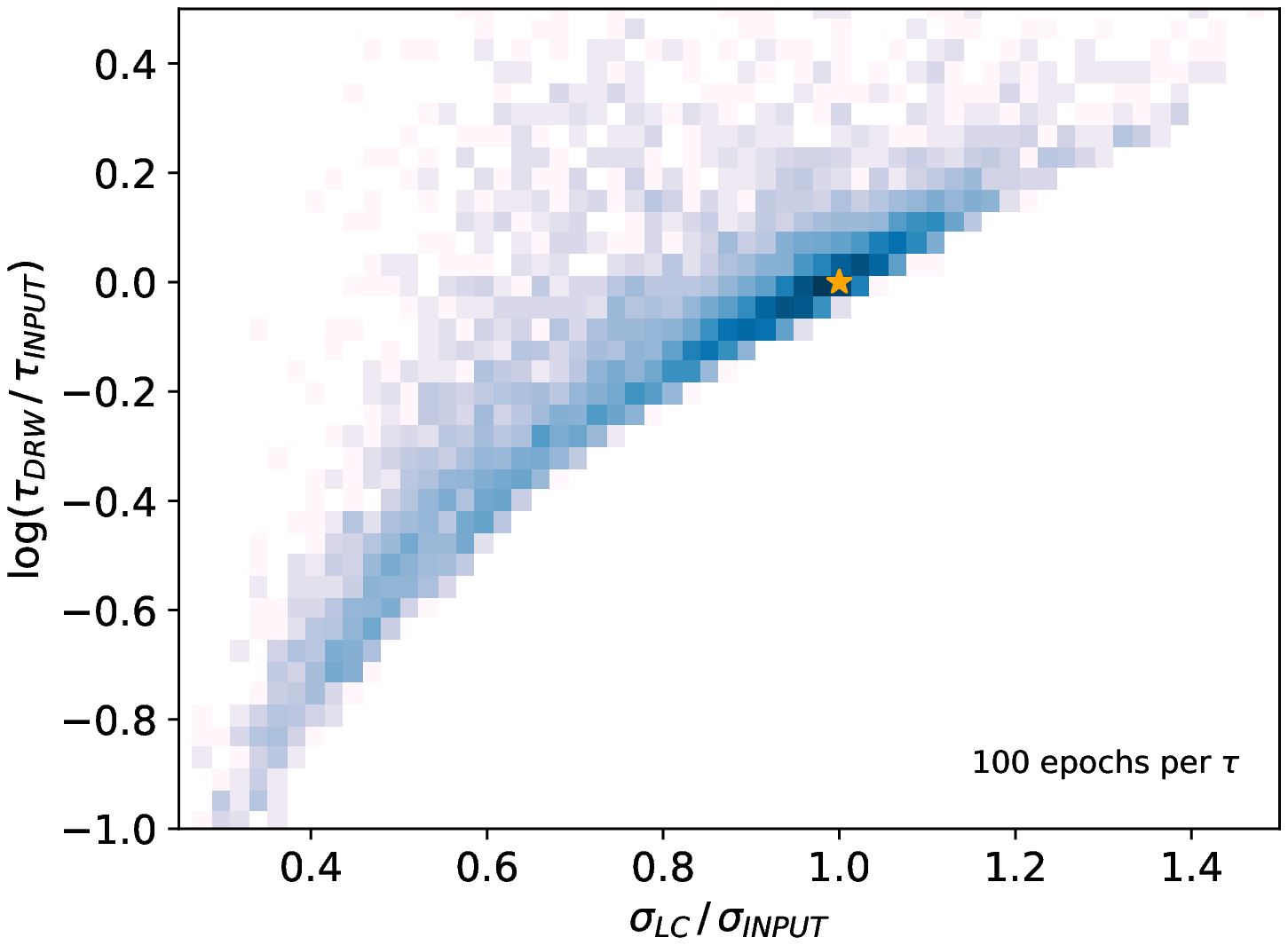}}

\FigCap{Output DRW parameters are shown. {\it Top panels:} The
  amplitudes obtained from DRW as a fraction of the input values are
  shown against the amplitude (dispersion) derived simply from light
  curves as a fraction of the input values. DRW delivers the amplitude
  in accordance with the amplitude present in the data. It is clear
  (from Fig.~2) that low dispersion fractions represent short light
  curves, meaning that DRW underestimates amplitudes for short light
  curves.  {\it Bottom panels:} The DRW time scales as a fraction of
  the input values are shown against the amplitude (dispersion)
  derived simply from light curves as a fraction of the input
  values. The shorter the light curve (the smaller the amplitude
  fractions) the more underestimated the DRW time scale as compared to
  the input value. The golden star marks the input value.}
\end{figure}

In Koz{\l}owski \etal (2010), we presented the dependence between the
variance in the continuous light curve and its length compared to the
decorrelation timescale $\tau$:
$${\rm var}(x)=\sigma^2\left[1-\frac{2}{x}+\frac{2}{x^2}
 \left(1-\exp(-x)\right)\right],\eqno(5)$$
where $x={\rm length}/\tau$ is the ratio of the survey duration to the
timescale $\tau$. This dependence is presented in the bottom panels of
Fig.~2 as the curved dotted line.

Next, we model these short light curves with the DRW model and measure
the two model parameters. We present the results in Figs.~3 and 4. In
top panels of Fig.~3, we show the ratio of the dispersion measured by
the DRW model to the input dispersion as a function of the ratio of
ordinary dispersion calculated from the light curves to the input
dispersion. We can see that DRW generally obtains dispersions that are
present in the data (as estimated by a simple dispersion calculation).
In the bottom panels of Fig.~3, we show the ratio of the time scale
measured with DRW to the input time scale as a function of the ratio
of ordinary dispersion calculated from the light curves to the input
dispersion. It is clear that the two parameters are correlated. The
smaller the dispersion the shorter the timescale obtained by DRW.

In Fig.~4, we present the ratio of the time scale measured with DRW to
the input time scale as a function of light curve length expressed in
time scales $\tau$. We can see that for light curve lengths longer
than about 30$\tau$ ($\log(\rm{length_{LC}}/\tau)=1.5$) the
DRW-measured time scales adequately represent the true value, while
for the shorter light curves the time scale seems to be
underestimated.
\begin{figure}[htb]
\vglue-3mm
\centerline{\includegraphics[width=6cm]{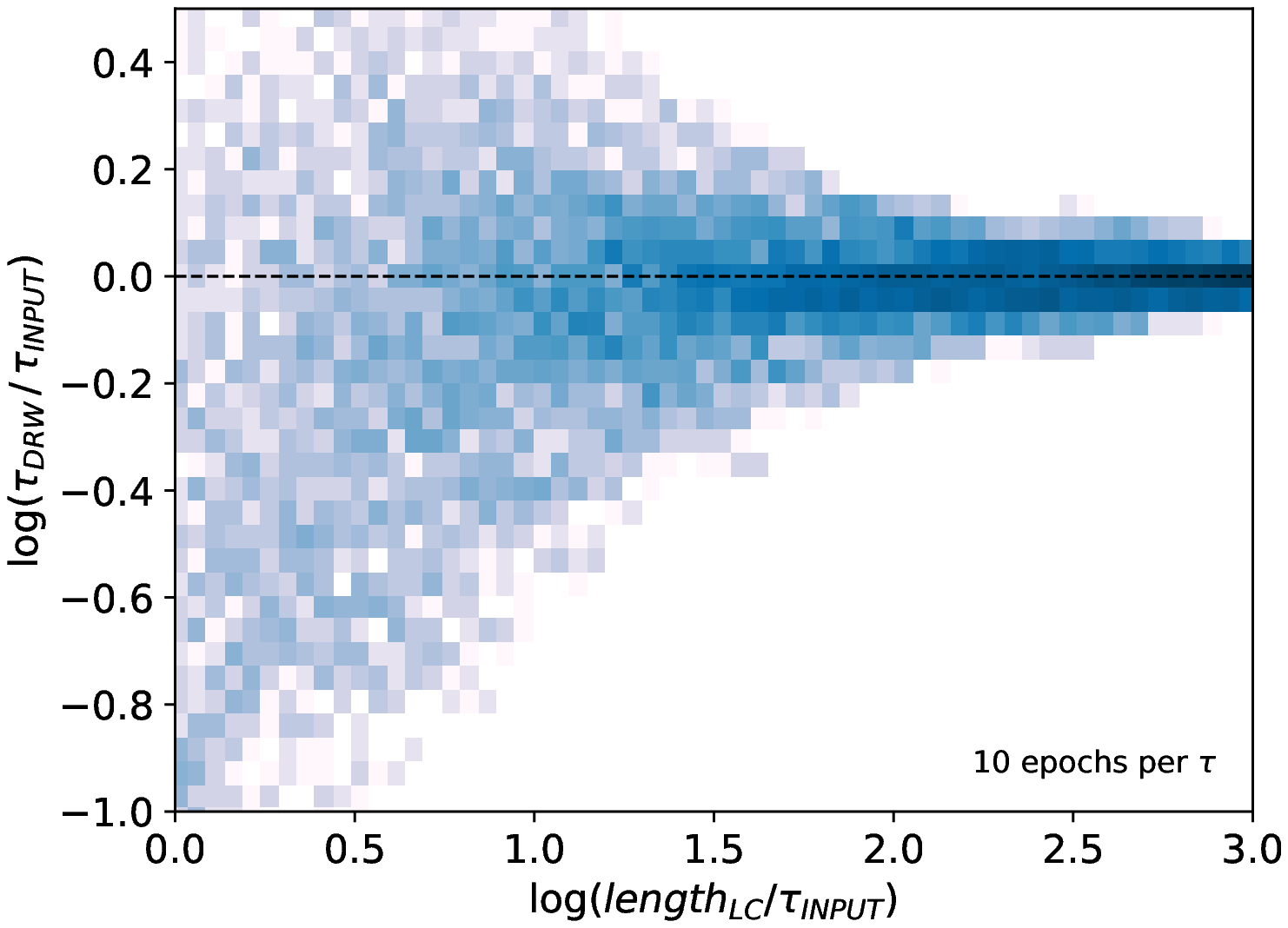}\includegraphics[width=6cm]{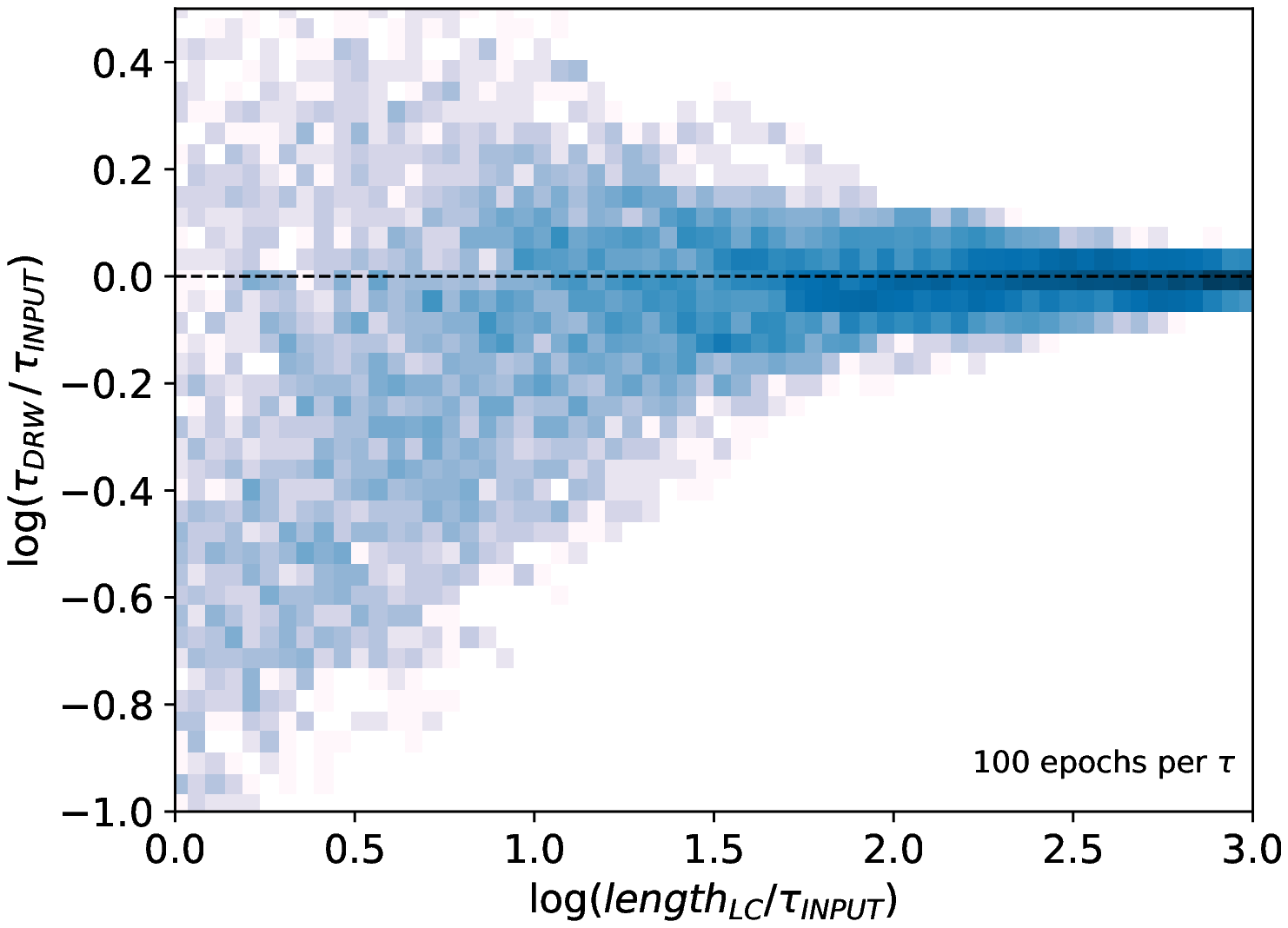}}
\FigCap{The DRW time scales as a fraction of the input values are
  shown against the light curve length as a fraction of the input time
  scale. The shorter the light curve, the more biased the measured DRW
  time scales -- typically toward shorter values. The length of a
  light curve must be at least 10 times the decorrelation time scale
  $\tau$ to reasonably recover the intrinsic process parameters.}
\end{figure}

In fact, it now appears that as early as Koz{\l}owski \etal (2010) paper,
we had all the ingredients to uncover the issues related to the data
length and their impact on the DRW model parameters. From that paper,
we knew that DRW is feasible to model both stochastic (DRW) and
deterministic data sets (periodic stars), while in Koz{\l}owski (2016b),
we found that DRW models stochastic processes with other exponential
covariance matrix equally well. From Koz{\l}owski \etal (2010), we knew
the variance of a light curve as a function of light curve length and
$\tau$ (Eq.~5) and we also commented that $\tau$ and $\sigma$ are
correlated.
\begin{figure}[b]
\hglue1pt{\includegraphics[width=6.7cm]{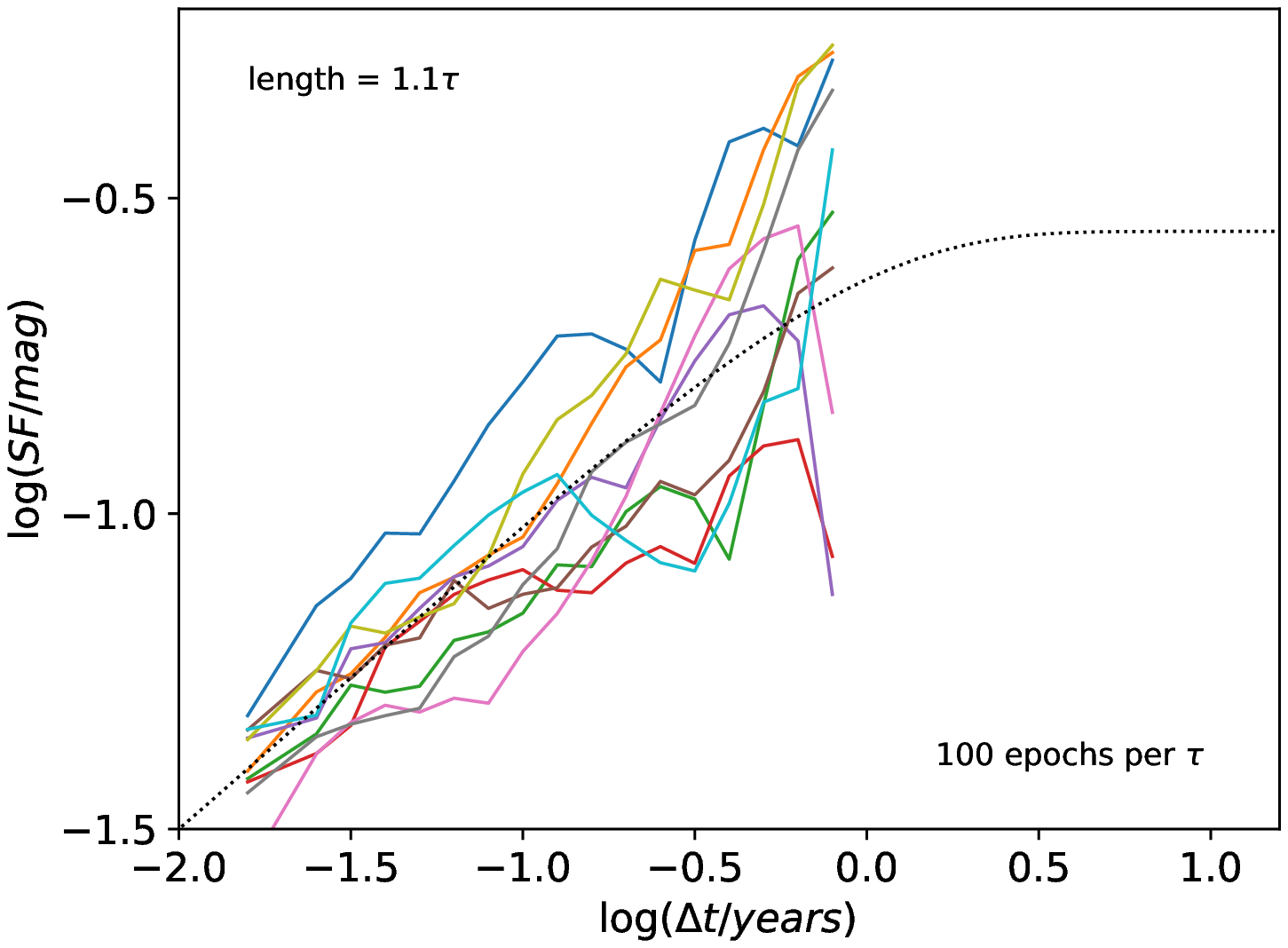}\includegraphics[width=6.7cm]{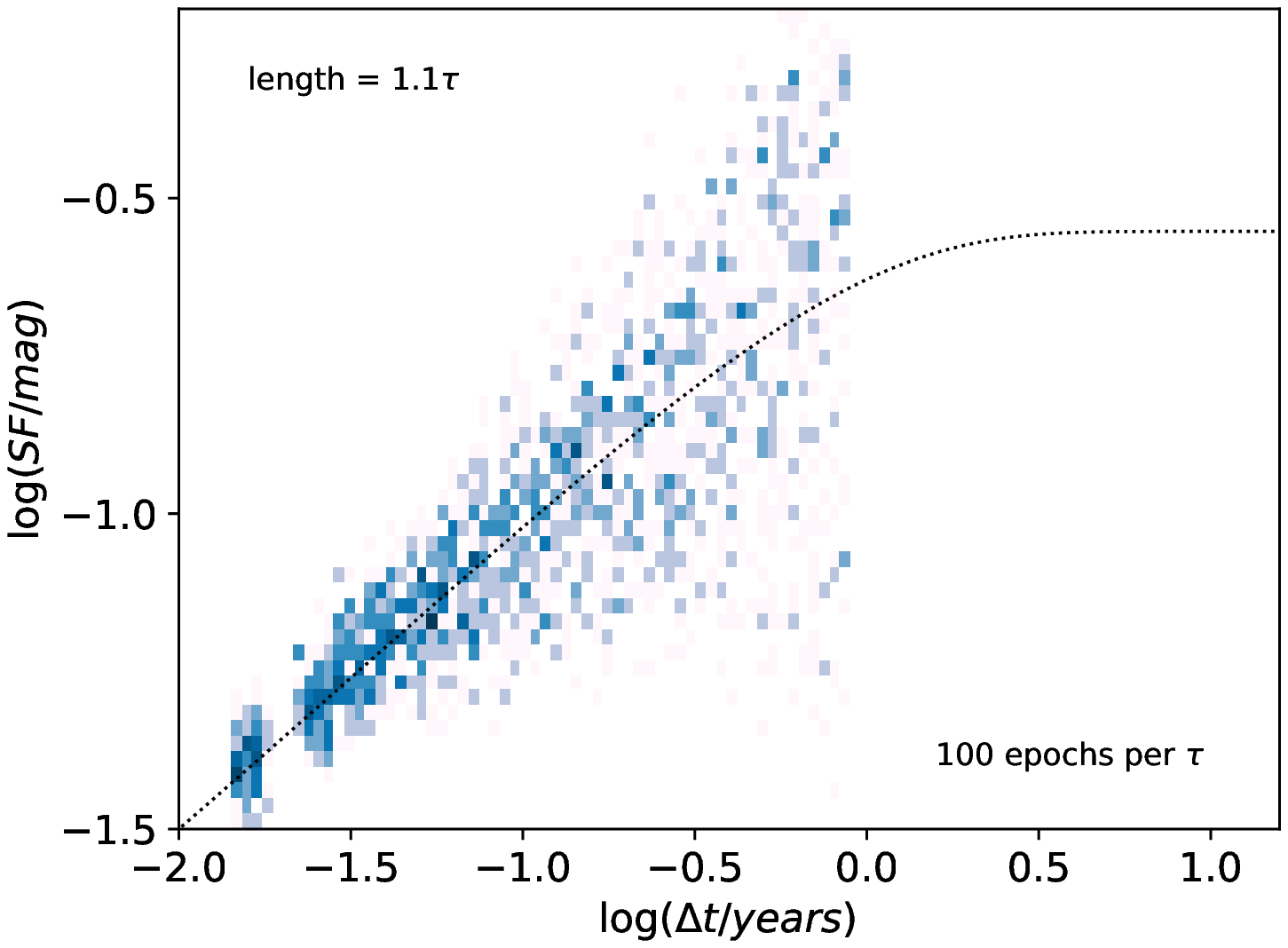}}
\vspace{0.3cm}
\hglue1pt{\includegraphics[width=6.7cm]{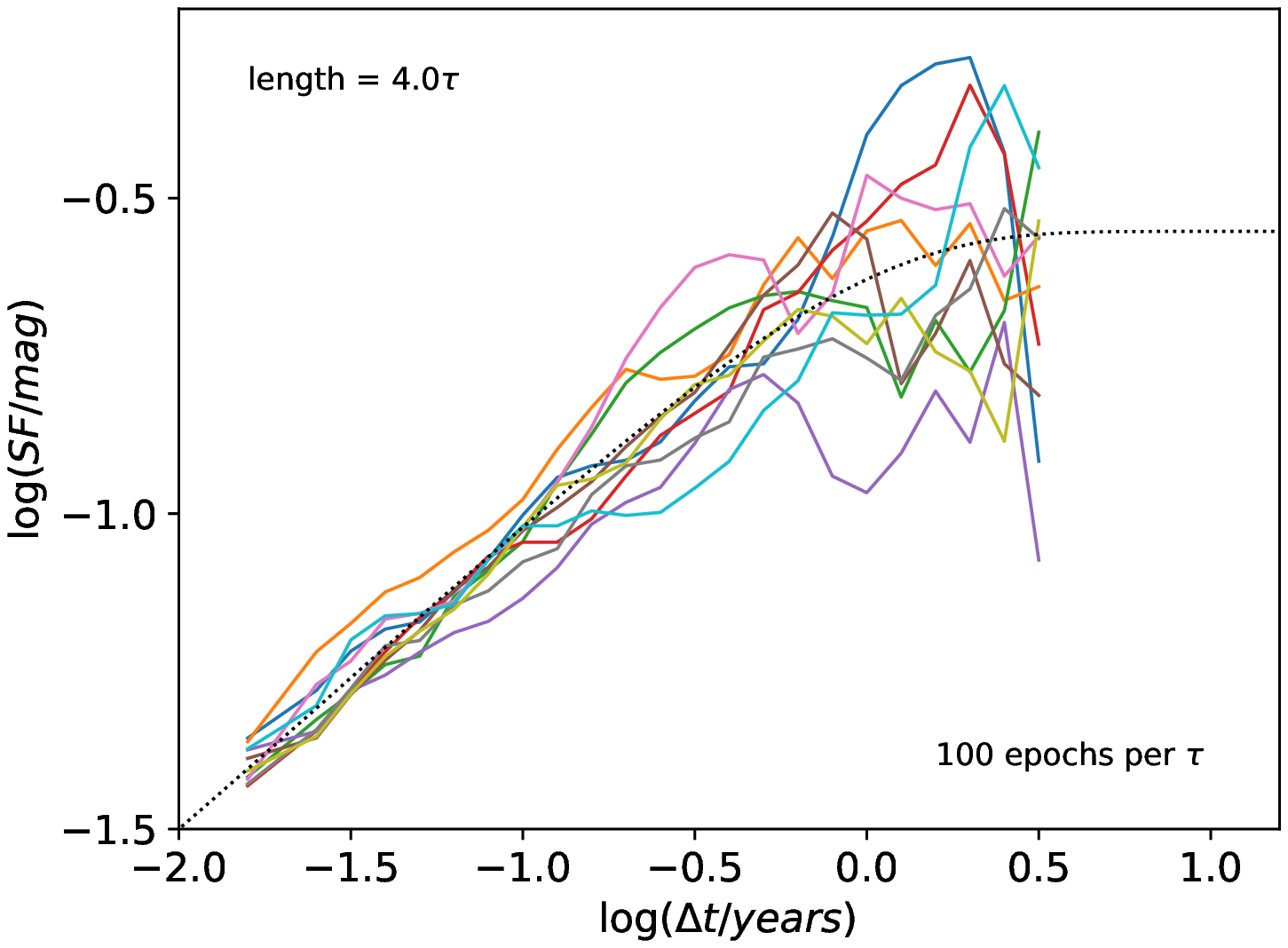}\includegraphics[width=6.7cm]{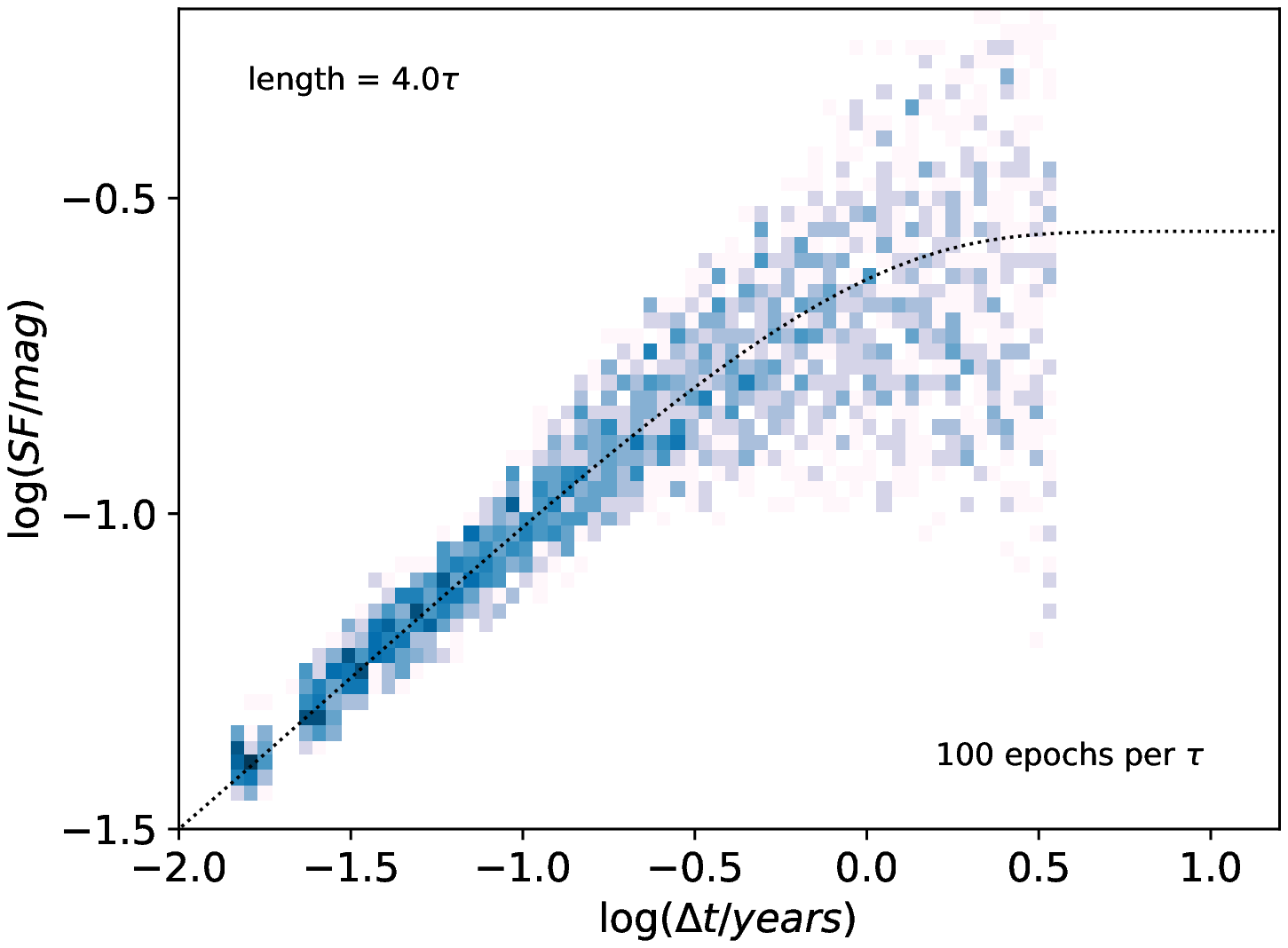}}
\vspace{0.3cm}
\hglue1pt{\includegraphics[width=6.7cm]{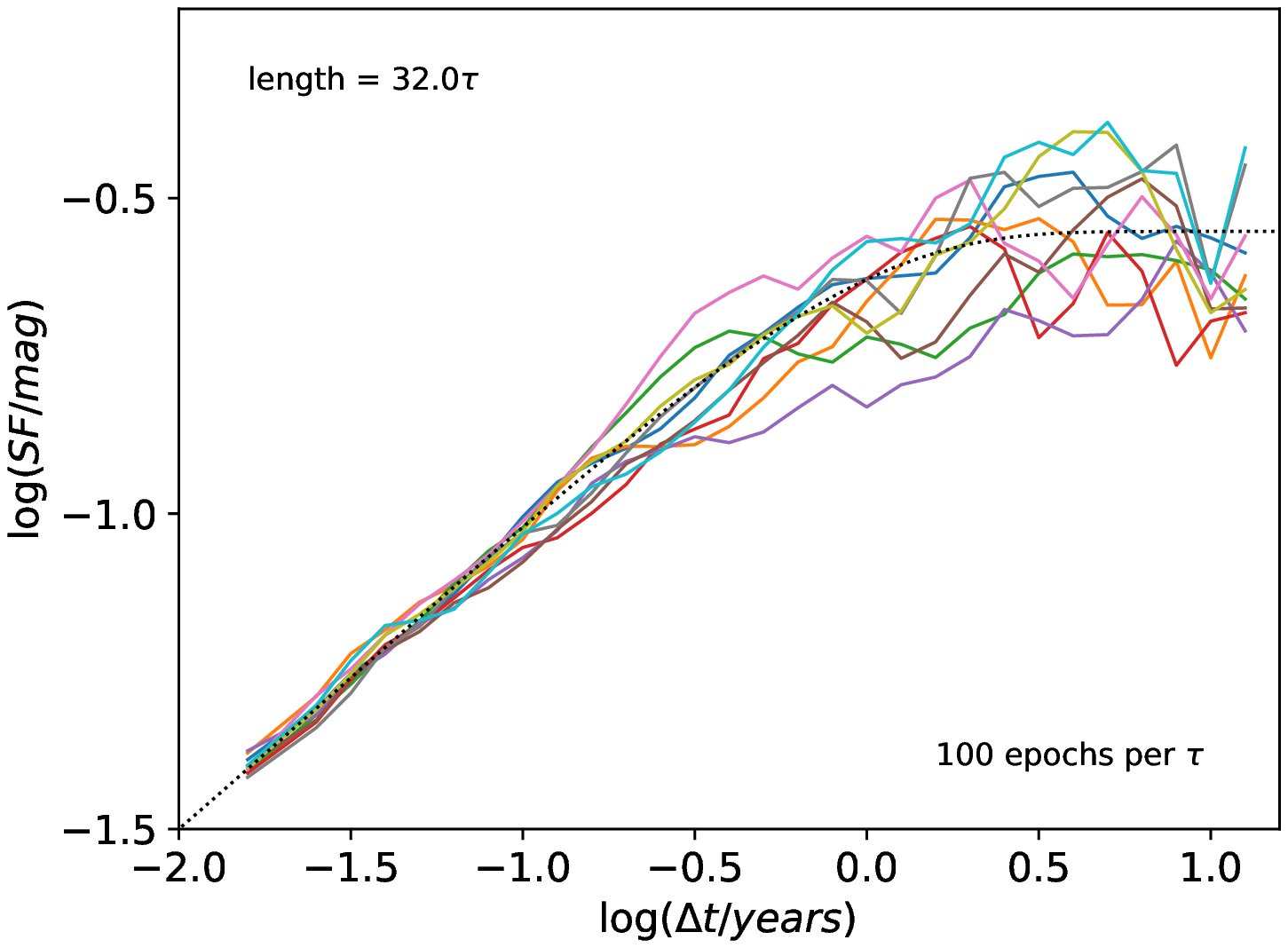}\includegraphics[width=6.7cm]{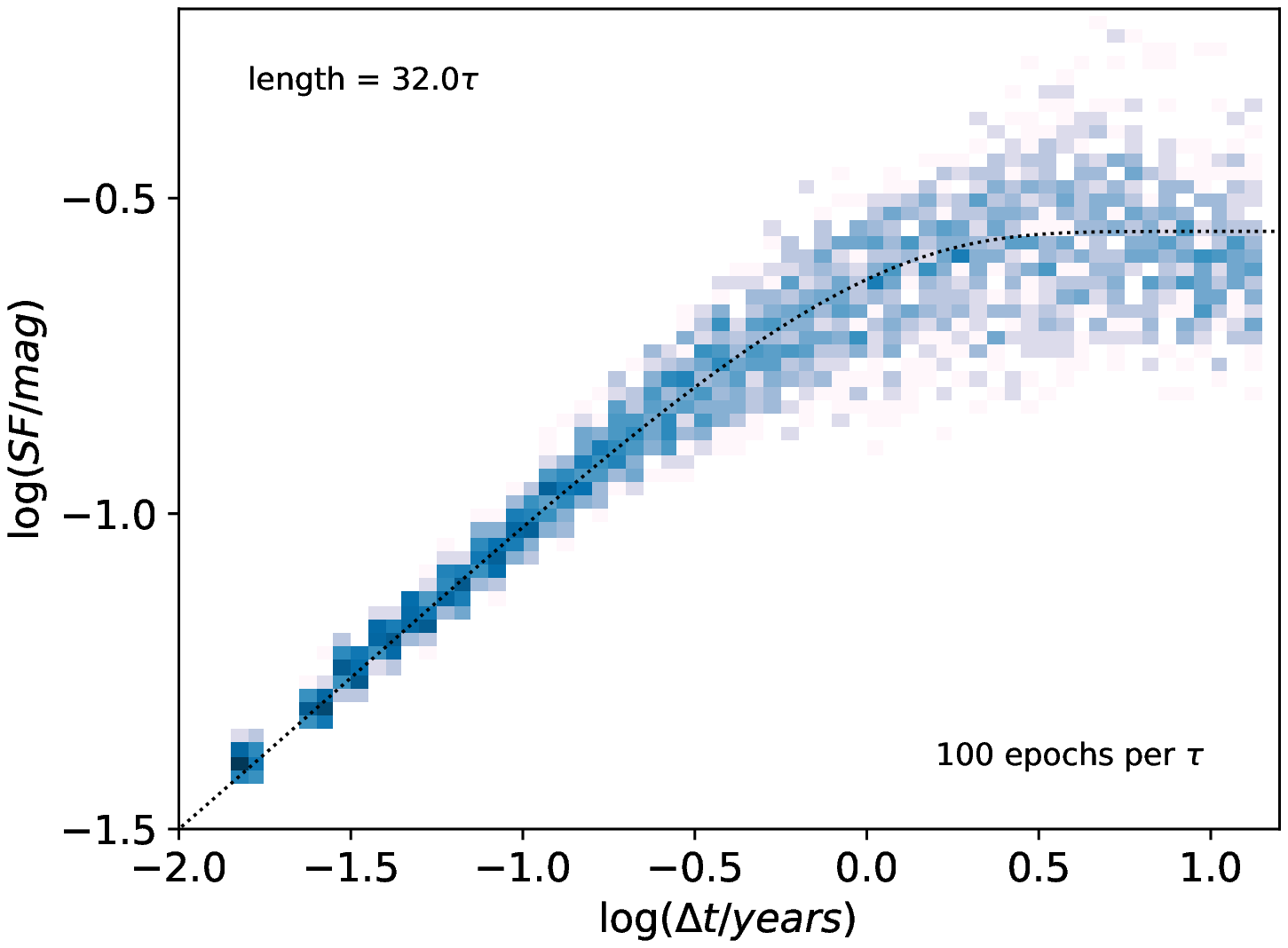}}
\vskip5pt
\FigCap{Structure functions for light curve lengths of $1.1\tau$ ({\it
    top row}), $4.0\tau$ ({\it middle row}), and $32.2\tau$
  ({\it bottom row}). The left column shows ten individual SFs,
  while the right column shows density histograms for 100 SFs. The
  dashed line is the intrinsic SF or the simulated process (and not a
  fit).}
\end{figure}

What can we learn from short light curves though? Since getting the
correct DRW parameters seems unlikely, we may try other methods to
uncover some information about variability. The key disadvantage of
DRW is that it has a fixed shape of the covariance matrix of the
signal.  It is designed in such a way that gives rise to a power
spectral distribution with the slope of $-2$ (${\rm PSD} \propto
\nu^{-2}$), the so called red noise at high frequencies. This is
reflected in the fixed slope of the structure function (SF)
$\gamma=0.5$ for ${S\!F}(\Delta t) \propto \Delta t^\gamma$, where
$\Delta t$ is the time difference between data points.

There exists some evidence that observed PSD and SF slopes for AGN
differ from the DRW values (\eg Mushotzky \etal 2011, Kasliwal,
Vogeley, and Richards 2017, Koz{\l}owski 2016a, Caplar, Lilly and
Trakhtenbrot 2017). In principle, the slope may depend on the physical
parameters of AGN (\eg Koz{\l}owski 2016a, Simm \etal 2016). Can we
measure these slopes from short light curves that do not represent a
stationary process?

To test this, we measured structure functions for short light curves
from our simulation (see a detailed elaboration on this topic in
Koz{\l}owski 2016a) and presented them in Fig.~5. The left column shows
10 individual SFs for very short ($1.1\tau$, top panel), short
($4.0\tau$, middle panel), and medium length ($32\tau$, bottom panel)
light curves taken from the original simulated high cadence light
curves. The right column shows the corresponding density histograms
based on 100 light curves. For the very short light curves, where the
light curve length is comparable to the decorrelation time scale, the
SFs are very noisy and obviously do not probe the bending
SF. Obtaining the SF slopes and drawing conclusions based on
individual SFs appears to be fruitless. Once the light curve length
grows the situation improves significantly. For light curves spanning
$32\tau$, we may be able to estimate correct slopes (at short time
scales) for individual objects. The full shape of SF can be measured
from many light curves representing the same process (the bottom-right
panel of Fig.~5). This is the so-called ``ensemble'' variability
measurement (\eg Vanden Berk \etal 2004, MacLeod \etal 2012, Vagnetti
\etal 2016, Koz{\l}owski 2017a, Li \etal 2018, Wang and Shi 2019). It
only works under assumption that AGNs with the same physical
parameters show the same variability properties.

\Section{Conclusions}
In this paper, we identified the origin of problems that one
encounters when modeling AGN light curves with DRW. By simulation
means, we showed that typical light curves -- that are of order of a
decade long with cadences of 3--30~d -- do not adequately represent
the underlying stochastic DRW process, assuming AGNs do indeed
generate DRW or DRW-like stochastic variability. Typical AGN light
curves do not have the properties of a stationary process, assuming
the intrinsic AGN variability is indeed due to such a
process. Therefore, it may be difficult, if not impossible, to
correctly reproduce the intrinsic process (to measure its parameters
with DRW) having the incomplete information about it.

In particular, we showed that the shorter the light curve the smaller
its variance as measured by standard procedure (Fig.~2). Then, we
identified a strong correlation between that dispersion and the one
measured from DRW modeling (Fig.~3). Since both DRW parameters are
correlated, increasingly smaller dispersions in increasingly shorter
light curves are reflected in increasingly shorter signal
decorrelation time scales as measured by DRW.

The DRW stochastic process is mathematically sound concept. As a
weakly stationary process, it requires its mean, variance, and
covariance (the auto-correlation function) to be time invariant. Once
the astronomical time-domain survey reach sufficient lengths to
fulfill the stationarity requirements, only then the measured
variability parameters will reflect the intrinsic ones.

\Acknow{S.K. acknowledges the financial support of the Polish National
  Science Center through the OPUS grant number 2018/31/B/ST9/00334.}

\end{document}